\documentclass{aastex} 
\usepackage{emulateapj5}
\shorttitle{X-ray Spectral variability of NGC~6251}
\shortauthors{Gliozzi et al.}
  \def\ngc{NGC~6251}

  \def\feka{Fe K$\alpha$}
   
  \def\xmm{{\it XMM-Newton}} 
  \def\asca{{\it ASCA}} 
   
  \def\rxte{{\it RXTE}} 
  \def\sax{{\it BeppoSAX}} 
   
  \def\rosat{{\it ROSAT}}

  \def\lum{erg s$^{-1}$}
  \def\flux{erg cm$^{-2}$ s$^{-1}$}

  \def\ltsima{$\; \buildrel < \over \sim \;$}
  \def\simlt{\lower.5ex\hbox{\ltsima}} 
  \def\gtsima{$\; \buildrel > \over \sim \;$}
  \def\simgt{\lower.5ex\hbox{\gtsima}} 

\begin{document}
\title{Long-Term X-Ray monitoring of NGC~6251:\\
Evidence for a jet-dominated radio galaxy}

\author{M. Gliozzi}
\affil{George Mason University, 4400 University Drive, Fairfax, VA 22030}

\author{I.E. Papadakis}
\affil{Physics Department, University of Crete, 710 03 Heraklion,
Crete, Greece}

\author{R.M. Sambruna}
\affil{NASA Goddard Space Flight Center, Code 661, Greenbelt, MD 20771}

\begin{abstract}
We present the first X-ray monitoring observations of the X-ray bright 
FR~I radio galaxy NGC~6251 observed with \rxte\ for 1 year. 
The primary goal of this study is to shed light on the origin of
the X-rays, by investigating the spectral variability with model-independent
methods coupled with time-resolved and flux-selected spectroscopy. 
The main results can be summarized as follows: 1) Throughout
the monitoring campaign, NGC~6251 was in relatively high-flux state
with an average 2--10 keV absorbed flux of the order of 4.5$\times 10^{-12}
~{\rm erg~cm^{-2}~s^{-1}}$ and a corresponding intrinsic luminosity of
6$\times 10^{42}~{\rm erg~s^{-1}}$.
2) The flux persistently changed with fluctuations of the order 
of $\sim$2 on time scales of 20-30 days.
3) When the hardness ratio is plotted against the average count rate, there
is evidence for a spectral hardening as the source brightens; this finding
is confirmed  by a flux-selected spectral analysis.
4) The fractional variability appears to be more pronounced in the hard
energy band (5--12 keV) than in the soft one (2.5--5 keV).
5) 2-month averaged and flux-limited energy spectra are adequately fitted 
by a power law.
A Fe K$\alpha$ line is never statistically required, although the presence of
a strong iron line cannot be ruled out, 
 due to the high upper limits on the
line equivalent width.
The inconsistency of the spectral variability behavior of \ngc\
with the typical trend observed in Seyfert galaxies and the 
similarity with blazars lead support to a jet-dominated scenario
during the RXTE monitoring campaign.
However, a possible contribution from a disk-corona system cannot be
ruled out.
\end{abstract}


\section{Introduction}
Non-blazar radio-loud active galactic nuclei (AGNs) --objects with jets forming
large viewing angles to the line of sight-- 
are traditionally divided into two classes:
Fanaroff-Riley II (FR~II), with 178 MHz powers $> 2
\times 10^{25}$ W/Hz and edge-darkened radio morphologies, and FR~I
galaxies with lower powers and more compact morphologies (Fanaroff \& Riley 
1974). 
For the same host galaxy optical
magnitude, FRIs produce about one order of magnitude less optical line
emission than FRIIs (Baum et al. 1995) and have fainter or
negligible UV continuum fluxes (Zirbel \& Baum 1995, Ho 1999).

While earlier models for the origin of the FRI/II dichotomy focused
mainly on accounting for their large-scale radio morphologies, more
recently new ideas have emerged concerning the nature of the central
engine in the two types of radio galaxies in an attempt to explain the
nuclear properties. One school of thought is that the nuclear X-ray
properties of FR~I and FR~II are related to a different accretion rate
onto the central supermassive black hole, with FRII being dominated by
relatively large values of $L/L_{Edd}$, while FRI would be accreting at
sub-Eddington rates, $L/L_{Edd} \ll 10^{-3}$ (e.g., Reynolds et
al. 1996; Ghisellini \& Celotti 2001). 
 
A fundamental step to gain insight into the nature of the central engine
in radio-loud AGNs is to understand whether the X-ray radiation is produced
by disk/corona systems as in Seyfert galaxies or by jets as in blazars.
While the time-averaged spectroscopy (due to
spectral degeneracy) and the pure temporal analysis (due to the fact
that both radio-quiet and radio-loud show strong variability) cannot
firmly discriminate between the two competing scenarios, their
combination, i.e., time-resolved spectral analysis and 
energy-selected temporal analysis, offers in principle a better way to 
distinguish between accretion-dominated and jet-dominated systems. 
This conclusion is supported by  
the strikingly different spectral variability behavior
shown by Seyfert-like objects (e.g., Papadakis et al. 2002; Markowitz \&
Edelson 2001) and by blazars (e.g., Zhang et al. 1999; 
Fossati et al. 2000; Gliozzi et al. 2006).

Here, we concentrate
on the X-ray nuclear properties of the nearby radio galaxy
NGC~6251 ($z$=0.024), which is a giant elliptical galaxy hosting a
supermassive black hole with mass $M_{\rm BH}\sim 4-8 \times10^8
M_{\odot}$ (Ferrarese \& Ford 1999). Based on its radio power at 
178 MHz, NGC~6251
is classified as an FR~I (e.g., Owen \& Laing 1989), whereas in the optical,
it is classified as a type-2 AGNs (e.g., Shuder \& Osterbrock
1981). 

Despite the intensive study of this source at all wavelengths,
the nature of the accretion process in NGC~6251 is still a matter of
debate. Based on the radio-to-X-ray spectral energy distribution, Ho
(1999) suggested that an Advection-Dominated Accretion Flow (ADAF) is
present in the nucleus of NGC~6251. On the other hand, Ferrarese \&
Ford (1999) and Melia et al. (2002) favored a standard accretion
disk. Finally, Mukherjee et al. (2002) and  Chiaberge et al. (2003) 
advocate a jet origin for the
broad-band emission, based on the possible association of \ngc\ with
the {\it EGRET} source 3EG~J1621+8203 and on
its spectral energy distribution, respectively. 

In the X-ray band, NGC~6251 has been previously
observed with various satellites. For example,
\rosat\ showed the presence of an unresolved nuclear source embedded in a
diffuse thermal emission with temperature $kT \sim 0.9$ keV,
associated to the galaxy's halo (Birkinshaw \& Worrall 1993). In
addition, the existence of a correlation between soft X-rays and radio
fluxes prompted Hardcastle \& Worrall (2000) to hypothesize a jet
origin for the soft X-rays. On the other hand, the prominent \feka\ line
(EW$\simeq$ 400--500 eV)
detected by \asca\ in 1994 (Turner et al. 1997; Sambruna et al. 1999)
suggested a standard Seyfert-like scenario with accretion-dominated
X-rays. \sax\ observed \ngc\ in July 2001 during a  high-flux state
(the 2--10 keV flux, $F_{\rm X}=4.7\times 10^{-12}$ \flux,
 was  $\sim$3 times larger than the value measured
by \asca\ 7 years earlier). Based on the absence of a prominent \feka\ line
(EW $<$ 100 eV) or other reprocessing features,
 Guainazzi et al. (2003) proposed a scenario
with two main spectral components: a blazar-like spectrum dominating
the high-flux state and a Seyfert-like spectrum emerging during the
low-flux state. More recently, in March 2002, \ngc\ was observed with 
\xmm\ when the 2--10 keV flux was $\sim$15\% lower than the \sax\
value. Despite the relatively high flux, the \xmm\ spectral results seem 
to support the
picture emerged from the \asca\ observation with the
presence of a prominent (EW$\sim$220 eV) and possibly broad ($\sigma>0.3$ keV)
\feka\ line (Gliozzi et al. 2004).
However, the existence of a broad \feka\ line in the \xmm\
spectrum is still a matter of debate (see Evans et al. 2005 and 
Gonz\'alez-Mart\'in et al. 2006 for a discording and a supporting view,
respectively).

The controversial results derived from previous X-ray studies 
highlight the impossibility of firmly determining the origin
of the X-rays in NGC~6251 based solely on time-averaged spectral
results. Here, we present the results from a systematic study of the 
long-term X-ray flux 
and spectral variability of \ngc\ using one year long proprietary Rossi X-ray
Timing Explorer (\rxte) observations in the 2-12 keV range.
We use model-independent methods and time-resolved spectroscopy to study 
the X-ray temporal and spectral properties of this source. 
The main purpose of this analysis is to shed light on the origin
of the X-rays and in particular on the role played by a jet in the X-rays.  
Once (if) the jet contribution is properly 
assessed, the physical parameters characterizing the accretion process onto the
supermassive black hole can be better constrained, and hence it is
possible to discriminate between competing theoretical models for the 
accretion process. 

The outline of the paper is as follows. In $\S~2$ we describe the
observations and data reduction. The main characteristics of the 
X-ray light curve are described in $\S~3$. In $\S~4$ we study the X-ray
spectral variability of \ngc\ with two model-independent methods. 
In $\S~5$ we describe the results of a time-resolved and flux-selected
spectral analyses. 
In $\S~6$ we discuss the results and their implications. 
In $\S~7$ we summarize the main results and conclusions.

\section{Observations and Data Reduction}

We use proprietary \rxte\ data of \ngc\ that was regularly observed for 
$\sim$ 1000 s once every 4 days between 2005 March 4  and 2006 February 27.  
The observations  were carried out
with the Proportional Counter Array (PCA; Jahoda et al. 1996),  and the
High-Energy X-Ray Timing Experiment (HEXTE; Rotschild et al. 1998) on \rxte.
Here we will consider only PCA data, because the signal-to-noise of the HEXTE
data is too low for a meaningful analysis.

The PCA  data were screened according to the following acceptance criteria: the
satellite was out of the South Atlantic Anomaly (SAA) for at least 30 minutes,
the Earth elevation angle was $\geq 10^{\circ}$, the offset from the nominal
optical position was $\leq 0^{\circ}\!\!.02$, and the parameter ELECTRON-2 was
$\leq 0.1$. The last criterion removes data with high particle background rates
in the Proportional Counter Units (PCUs). The PCA background spectra and light
curves were determined using the ${\rm L}7-240$ model developed at the \rxte\
Guest Observer Facility (GOF) and implemented by the program {\tt pcabackest}
v.3.0.   This model is appropriate for ``faint'' sources, i.e., those
producing count rates less than 40 ${\rm s^{-1}~PCU^{-1}}$. 

All the above tasks were carried out using the {\tt FTOOLS} v.6.2 software
package and with the help of the  \verb+REX+  script provided by the \rxte\
GOF. Data were initially extracted with 16 s time resolution and subsequently
re-binned at different bin widths depending on the application.  The current
temporal analysis is restricted to PCA, STANDARD-2 mode, 2--12.5 keV, Layer 1
data, because that is where the PCA is best calibrated and most sensitive. PCUs
0 and 2 were turned on throughout the monitoring campaign. However, since the
propane layer on PCU0 was damaged in May 2000, causing  a systematic increase
of the background, we conservatively use only PCU2 for our analysis (see
below). All quoted count rates are therefore for one PCU.

The spectral analysis of PCA data was performed using the {\tt XSPEC v.12.3.1}
software package (Arnaud 1996). We used PCA response matrices and effective
area curves created specifically for the individual observations by the program
{\tt pcarsp}, taking into account the evolution of the detector properties. 
All the spectra were re-binned so that each bin contained enough counts for the
$\chi^2$ statistic to be valid. Fits were performed in the energy range 
2.5--12.5 keV, where the signal-to-noise ratio is the highest.

\section{The X-ray Light Curves}
Although \ngc\ is generally considered an X-ray bright source --  its
average flux is of the order of $4\times 10^{-12}$ \flux\  with a corresponding
luminosity L$_{\rm 2-10~keV} \sim 5\times10^{42}$ \lum\ that is nearly a
factor 4 
larger than the typical values observed in low-power radio galaxies 
(Donato et al. 2004) -- it is rather weak for the \rxte\ capabilities.
Therefore, before starting a detailed analysis of temporal 
properties, it is necessary to demonstrate that the variability observed  
cannot be ascribed to uncertainties in the \rxte\ background  or to other
artifacts. To this end, we have performed the following
test: We have compared the background-subtracted light curves obtained
using PCU2 layer 1 and PCU2 layer 3. Since the genuine signal in layer
3 is quite small, its light curve can be used as a proxy to check how
well the background model works. If the latter light curve is
significantly variable with a pattern similar to the one produced
using layer 1, then the variability is simply due to un-modeled
variations of the background. Conversely, if the PCU2 layer 3 light
curve does not show any pronounced variability or if the flux changes
are uncorrelated with those observed in the layer 1 light curve, we
can safely conclude that the variability detected in \ngc\
is real.  

\begin{figure}
\begin{center}
\includegraphics[bb=85 15 475 405,clip=,angle=0,width=9cm]{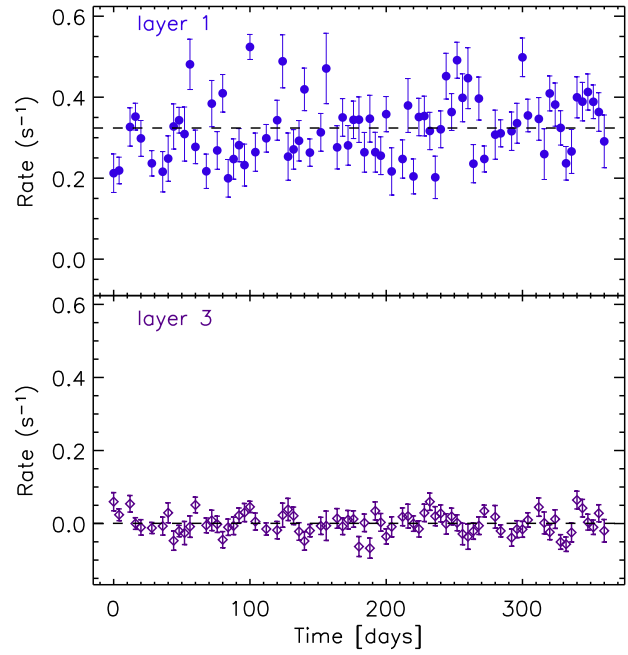}
\end{center}
\caption{Top panel: \rxte\ PCA light curve in the 2--10 keV energy
band, using PCU2 layer 1. Bottom panel: PCU2 layer 3 light curve.
Time bins are 4 days.  The dashed lines are the average count rate
level. 
}
\label{figure:fig1}
\end{figure}

\begin{figure}
\begin{center}
\includegraphics[bb=30 30 360 300,clip=,angle=0,width=9cm]{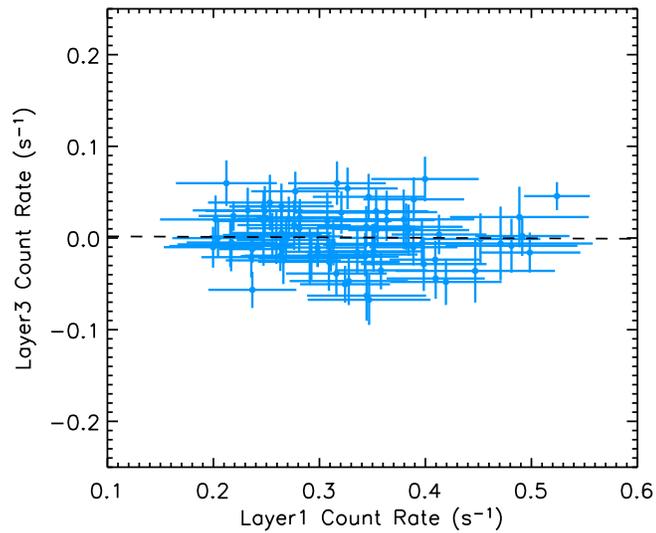}
\end{center}
\caption{Layer 3 plotted versus the layer 1 count rate.
The dashed line represents the best linear fit.}
\label{figure:fig2}
\end{figure}

The two light curves in the 2--10 keV range (where the
background PCA model is better parameterized; see Jahoda et al. 2006
for more details) are shown in Figure~\ref{figure:fig1}. A visual inspection
of this figure suggests that the variability in the layer 1 light curve is 
much  
more pronounced than the one observed in layer 3. Indeed, the mean count rate
of layer 3 is consistent with zero, indicating that on long timescales
the background model works adequately. However, statistically speaking both 
light curves are considered variable: $\chi^2=$ 229.5 (layer 1) and 128.3 
(layer 3) for 78 degrees of freedom (hereafter dof). On the other hand,
a formal analysis based on the excess variance (i.e. the variance corrected 
for statistical errors), indicates that variability associated with the 
layer 1 light curve is more than one order of magnitude larger than that 
associated with layer 3 ($\sigma_{\rm xs}= 4\times10^{-3}$ and 
$3\times10^{-4}~
{\rm s^{-2}}$, respectively). Further support to the fact that the 
variability associated with layer 1 is genuine comes from a correlation 
analysis: When layer 3 is plotted versus the layer 1 count rate (see 
Figure~\ref{figure:fig2}), no correlation is observed, as confirmed by 
a least square linear fit analysis, $y=0.002(\pm0.015)-0.005(\pm0.048)x$.
We therefore conclude that most of the count rate changes observed in layer 1
are associated with genuine intrinsic variations of the X-ray source in \ngc.

\subsection{PCU0 versus PCU2}
\begin{figure}
\begin{center}
\includegraphics[bb=100 25 430 460,clip=,angle=0,width=12cm]{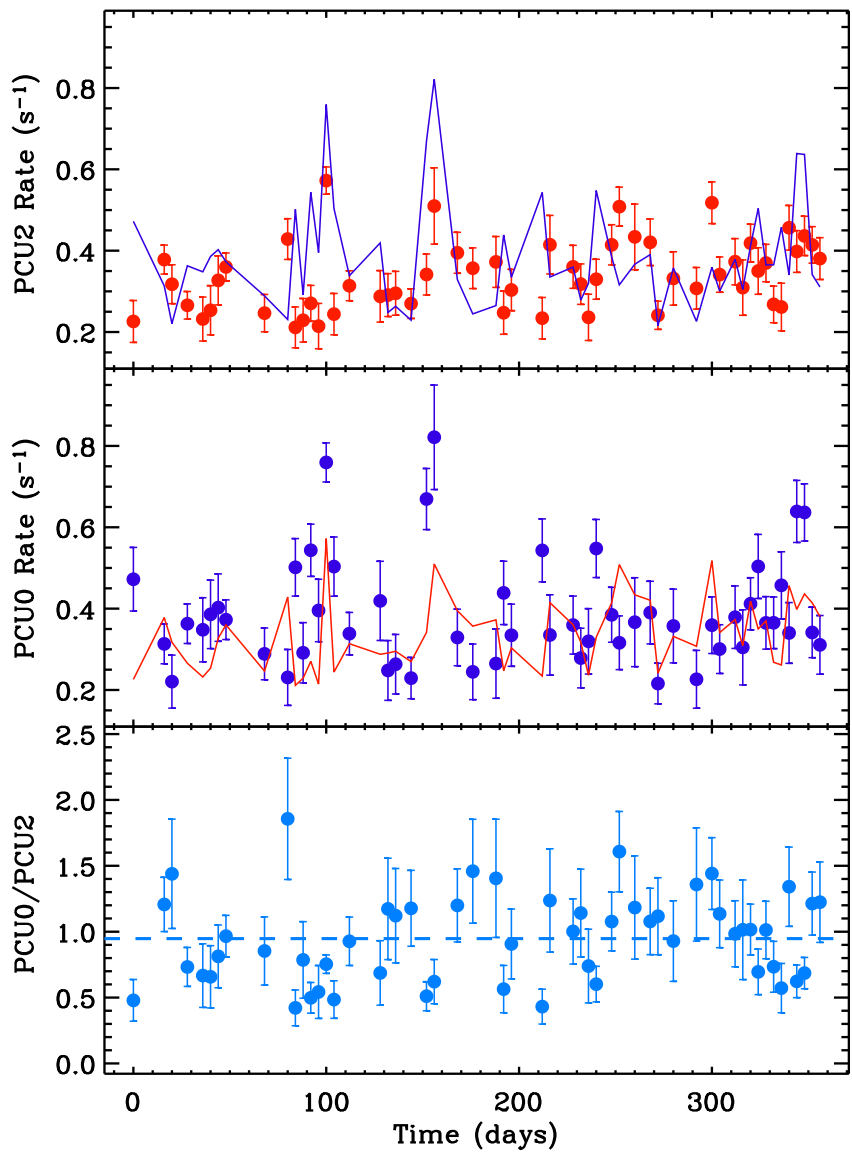}
\end{center}
\caption{Top panel: PCU2 light curve of \ngc\ in the 2--10 keV range; the 
solid line represents the PCU0 light curve.
Middle panel: PCU0 light curve of \ngc\ in the 2--12 keV range; the 
solid line represents the PCU2 light curve.
Bottom panel: PCU0/PCU2 light curve; the dashed line represents the average 
value of the ratio PCU0/PCU2. Time bins are 4 days.}
\label{figure:fig3}
\end{figure}

In order to maximize the signal-to-noise (S/N) ratio of the light curves, 
one can combine the 2 PCUs at work during
the monitoring campaign (i.e., PCU0 and PCU2), provided that they are
consistent with each other. In particular, it is  necessary to test
whether PCU0, partially damaged since May 2000, is compatible with PCU2.
Figure~\ref{figure:fig3} shows the background-subtracted, 2--12 keV  light
curves of PCU2 (top panel), PCU0 (middle panel), each with superimposed the
light curve of the other PCU (represented by continuous lines in top and 
middle panels), and their ratio (bottom panel)
with a time bin of 4 days. A visual inspection of  this
figure  suggests that the PCU0 and PCU2 light curves are not fully consistent
with each other.
Indeed, a formal check based on a $\chi^2$ test  indicates that the ratio
PCU0/PCU2 is not consistent  with the hypothesis of constancy  ($\chi^2=160.5$
for 54 dof). Since the propane layer on PCU0 was
damaged a few years ago, we decided to work with the PCU2 data only. For
completeness, we have also performed the same data analysis on the PCU0 data, 
and the results are in general agreement with those from the PCU2. 

Whereas for the temporal analysis the PCU0 data need to be necessarily 
excluded because they appear to be inconsistent with the PCU2 data, PCU0
spectral data should not be a priori ruled out for an analysis averaged over 
two-month periods. Unfortunately, also the 2-month averaged PCU0 spectra
appear to be at odds with the PCU2 spectral results, with unphysically large
changes for the photon index (with a seesaw trend between $\Gamma\sim 2$ and
 $\Gamma\sim 3$) and very large uncertainties. 
Therefore, also for the spectral analysis (see $\S5$) we use only PCU2 data.

\begin{figure}
\begin{center}
\includegraphics[bb=105 25 425 455,clip=,angle=0,width=12.cm]{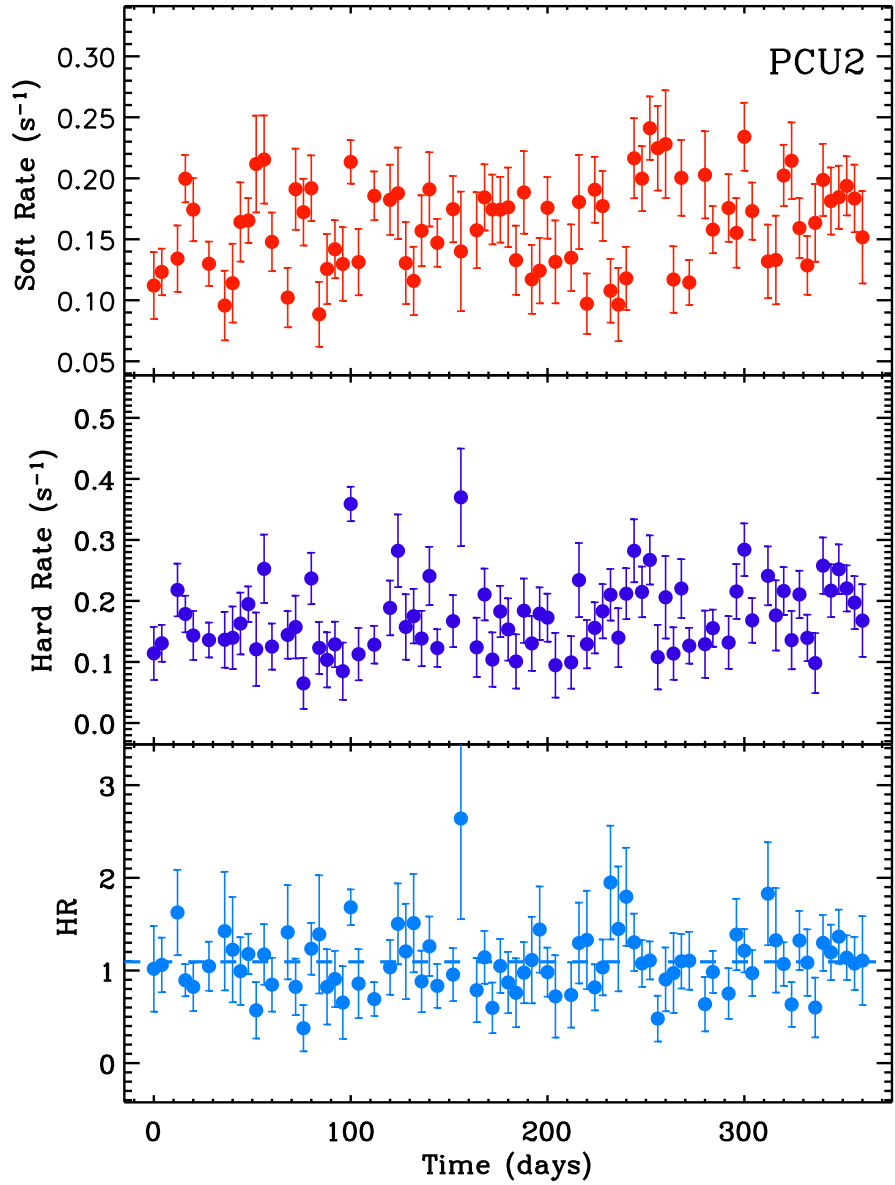}
\caption{Top panel: Soft (2.5--5 keV) light curve of \ngc\ using PCU2 data only.
Middle panel: Hard (5--12 keV) light curve.
Bottom panel: Hardness Hard/Soft ratio light curve; the solid line represents the 
average value of the hardness ratio. Time bins are 4 days.} 
\label{figure:fig4}
\end{center}
\end{figure}

Using data from PCU2 only, we constructed background subtracted light curves 
in two energy bands, namely the ``soft" (2.5--5 keV) and the ``hard" 
(5--12 keV) band.  We
show them in Figure~\ref{figure:fig4}, together with the  hardness ratio
(hard/soft) plotted versus time.
Time bins are 4 days. The soft and hard band light curves  are
significantly variable ($\chi^2=113.1/78$ and $\chi^2=175.6/78$, respectively)
and appear to be qualitatively similar. This
is formally confirmed by the hardness ratio light curve, which appears to be 
constant ($\chi^2=70.7/78$). 

We conclude
that the flux variations observed in \ngc\
on timescales of weeks-months are not associated with spectral variations. 

\section{Spectral variability: model-independent analysis}
For this study we use simple methods such as hardness ratio versus count rate 
plots and the fractional variability versus energy plots. These can provide 
useful
information without  any a priori assumption regarding the shape of the X-ray
continuum spectrum. Thus, the results from the study of these plots
can be considered as ``model-independent". 

\begin{figure}
\begin{center}
\includegraphics[bb=45 32 355 300,clip=,angle=0,width=9.cm]{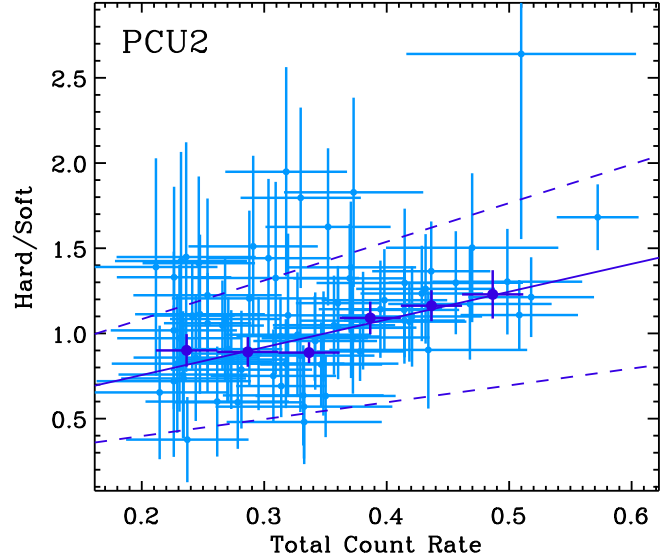}
\caption{Hardness ratio (5--15 keV/2.5--5 keV) plotted versus the total count 
rate. The gray (light blue in color) small symbols correspond to the un-binned
data points, whereas the darker, larger symbols refer to the weighted mean of
data points in bins of fixed size of 0.05 c/s along the x-axis.
The continuous line represents the best linear fit from the binned data,
whereas the dashed lines are the $\pm 1\sigma$ departure from the best fit. }
\label{figure:fig5}
\end{center}
\end{figure}

Figure~\ref{figure:fig5} shows the Hard/Soft X-ray color 
(5--12 keV)/(2.5--5 keV) plotted versus the
total count rate (2.5--12 keV) for un-binned (smaller, gray symbols) and 
binned data (larger, darker symbols), respectively. 
A visual inspection of this figure suggests the presence of a positive
trend between $HR$ and the count rate (i.e., the spectrum hardens
as the flux increases), although within a large scatter. 
This positive trend is apparently confirmed by the linear fit of the data,
$y=(0.2\pm0.2)+(2.3\pm0.5)x$ with $\chi^2/dof=43.9/77$, which suggests
the presence of a positive correlation at 4.6$\sigma$ confidence level. This
analysis was performed
using the routine \verb+fitexy+ (Press et al. 1997) that accounts 
for the errors not only on the y-axis but along the x-axis as well. 

Better insight into the presence of correlations between $HR$ and total count
rate can be obtained by investigating the binned data. To this end, we have
binned the data, shown in Fig.~\ref{figure:fig5}, using count rate bins of 
fixed size (0.05 counts/s), and
computed the weighted mean (along both the y and x-axis) of all the points
which fall into a bin. The error on the weighted mean is computed
following Bevington's prescriptions (Bevington 1969). Only bins with at least
5 data points have been plotted in Figure~\ref{figure:fig5} and considered
for the linear fit. In this case, the best linear fit, 
$y=(0.4\pm0.2)+(1.6\pm0.6)$ with $\chi^2/dof=1.44/4$, (shown in 
Fig.~\ref{figure:fig5} with its 
1$\sigma$ uncertainties), indicates that,
if a few outliers with large count rate (and error-bar) are neglected, 
the significance of a positive correlation reduces to 2.67$\sigma$. This is
slightly lower than the 3$\sigma$ level, which is generally accepted for
significant variations, but it still implies a positive correlation at
a confidence level of $\sim$ 99\%.

Another simple way to quantify the variability properties of \ngc\, without
considering the time ordering of the values in the light curves, is based on
the fractional variability parameter $F_{\rm var}$.  This is a commonly 
used measure of the intrinsic
variability amplitude relative to the mean count rate, corrected for the effect
of random errors, i.e., \begin{equation} F_{\rm
var}={(\sigma^2-\Delta^2)^{1/2}\over\langle r\rangle} \end{equation} 
\noindent
where $\sigma^2$ is the variance, $\langle r\rangle$ the unweighted mean count
rate, and $\Delta^2$ the mean square value of the uncertainties associated with
each individual count rate. The error on $F_{\rm var}$ has been estimated following Vaughan et al. (2003).
We computed $F_{\rm var}$ on selected energy bands, with
mean count rates similar and sufficiently
high. 

\begin{figure}
\centering
\includegraphics[bb=65 35 345 290,clip=,angle=0,width=9.cm]{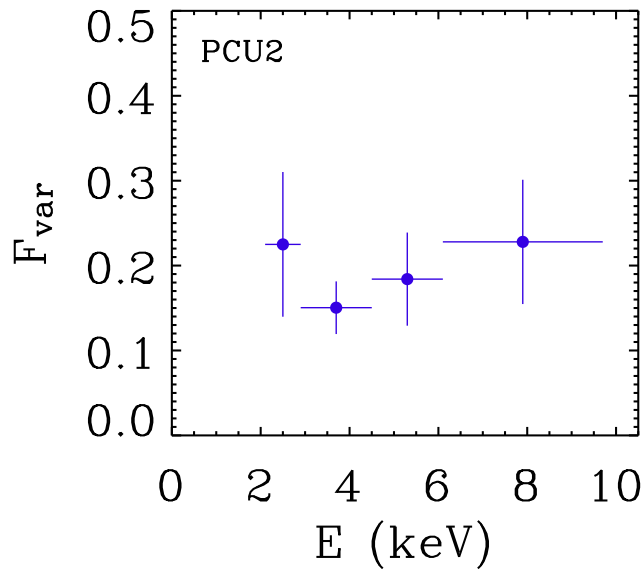}
\caption{Fractional variability amplitude as a function of
energy for \ngc. 
The error-bars along the x axis simply represent
the energy band width. The error bars along the y axis are computed following
Vaughan et al. 2003.
} 
\label{figure:fig6}
\end{figure}

The results are plotted in Figure~\ref{figure:fig6} and
suggest that, over the 3--8 keV energy range, a weak positive trend seems
to be present. However, the large uncertainties on $F_{\rm var}$ due to the 
low count rate in the narrow energy-selected bands hamper this kind of 
analysis. Indeed, statistically speaking, the trend shown in 
Figure~\ref{figure:fig6} is consistent with the hypothesis of $F_{\rm var}$
being constant in the energy range probed by \rxte. If only two
broader energy bands are used for this analysis, the resulting fractional
variability in the hard energy band 5--12 keV, $F_{\rm var,hard}=(20\pm3)\%$
appears to be larger than the value, $F_{\rm var,soft}=(10\pm5)\%$, obtained
in the soft band 2.5--5 keV. However, the $F_{\rm var,hard}-
F_{\rm var,soft}=(10\pm6)\%$ difference is significant at the 1.7$\sigma$ level
only.

In summary, the results from the study of the  $HR-ct$ plots show 
evidence for a  positive correlation between hardness ratio and
total count rate.  Similarly, an analysis of the fractional variability
suggests that $F_{\rm var}$ is more pronounced in the hard energy band
than in the soft one.
However, due to the rather limited signal-to-noise ratio of our data these
results are significant at just the $2.7$ and $1.7\sigma$ level, respectively.

\section{Spectral Analysis}

\subsection{Time-resolved Spectroscopy}
Given that the data consist of short snapshots spanning a long temporal 
baseline,
they are in principle well suited for monitoring the spectral variability of 
\ngc. However, due to the limited S/N of the data, the spectral slope 
$\Gamma$, measured
from spectra of individual observations, cannot be adequately constrained
and hence the spectral variability cannot be investigated in such a way. Indeed,
if we plot the values of $\Gamma$ versus time, no variations are detected due 
to the large errors on  $\Gamma$. This is fully consistent with the results
from the $HR$-time plot described in $\S4$. In order to increase the S/N  
and investigate the presence of possible spectral variations, we use spectra
averaged over two-month intervals. This
choice is a trade-off between the necessity of  accumulate sufficient counts
for a reliable spectral analysis and the need to use limited temporal intervals
to minimize the effects of the slow drift in the detector gain.

\begin{figure}
\includegraphics[bb=65 5 575 715,clip=,angle=-90,width=16.cm]{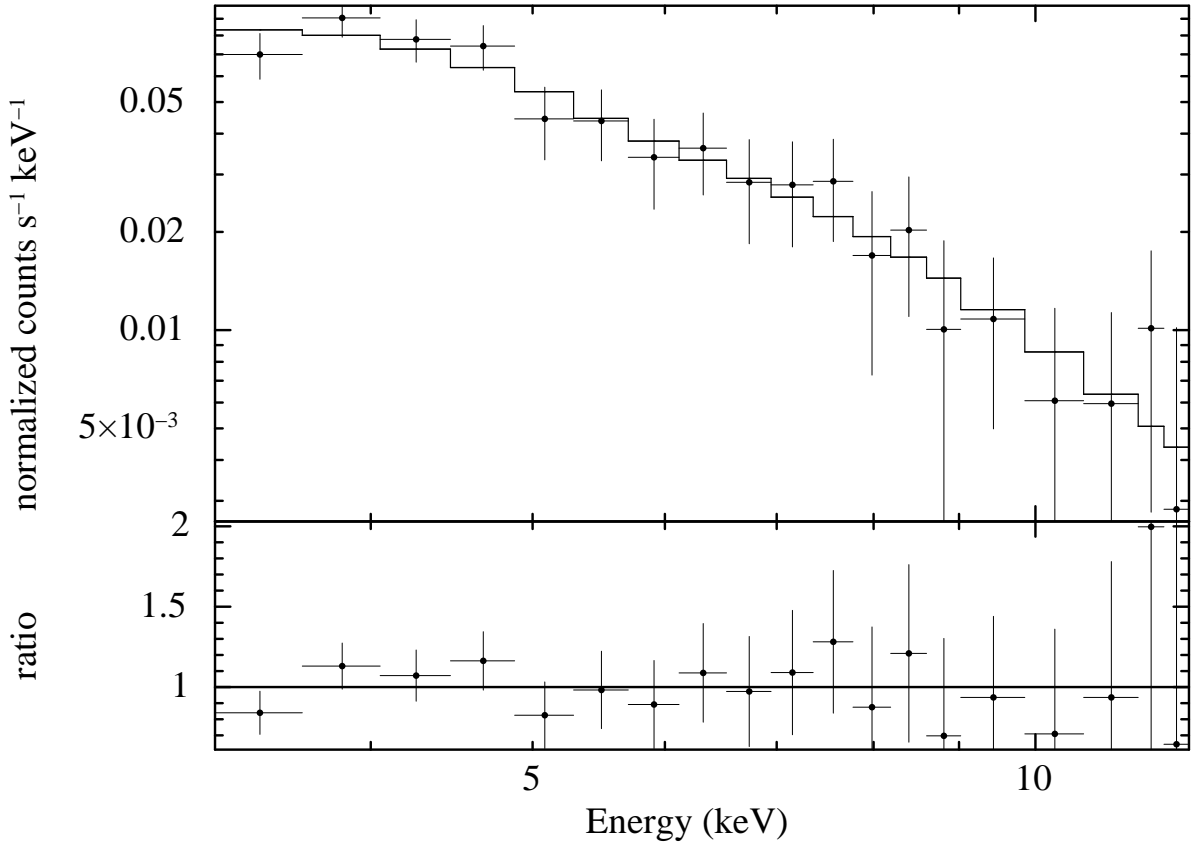}
\caption{PCA spectrum of \ngc\ during the third 2-month interval obtained 
using PCU2 data only. The model used is a simple power law absorbed by
Galactic $N_{\rm H}$.} 
\label{figure:fig7}
\end{figure}

We fitted each two-month spectrum with a simple power-law (PL) model 
absorbed by Galactic $N_{\rm H}$ ($5.65\times10^{20}{\rm~cm^{-2}}$). 
The model fits all the data reasonably well,  as indicated by 
Fig.~\ref{figure:fig7} that shows a typical spectrum fitted with a
simple power law.
The best-fit results are listed in Table 1
and can be summarized as follows. A simple PL model provides
an acceptable parametrization for all spectra. The photon indices are all
rather steep ($\Gamma\sim2.5$) and consistent with each other within the 
errors. In other words, our results suggest that the source's spectrum
does not vary significantly on timescales longer than two months. The
weighted spectral slope mean is $2.5\pm0.1$.
Adding a Gaussian line 
at 6.4 keV to the PL continuum model  does not improve the fit significantly 
in any of the six spectra, but the 90\% confidence upper limits on the 
equivalent
width are relatively high (EW$\sim$200-700 eV). 

In order to better constrain  $\Gamma$ and the line EW,
we have tried a comparison between the first and the last 6 months
of monitoring campaign, by fitting together 3 two-month spectra at a time.
The results -- $F_{\rm 2-10~keV}=4.3\times 10^{-12}$ \flux, 
$\Gamma=2.53\pm0.18$, $EW < 212$ eV during the first half, and 
 $F_{\rm 2-10~keV}=4.5\times 10^{-12}$ \flux, 
$\Gamma=2.46\pm0.17$, $EW < 144$ eV during the second half --
are fully consistent with each other.

Since there are no indications for long-term spectral variability, we
have fitted the 6 two-month spectra together. 
This yielded $\Gamma=2.5\pm0.1$ and a significantly smaller upper limit
on the line strength:
EW$<$104 eV. In the 2--10 keV energy
band, we obtained
an average absorbed flux of 4.4$\times 10^{-12}$ \flux, and a corresponding
intrinsic luminosity  of  $6.3\times 10^{42}$ \lum,
assuming $H_0=71{\rm~km~s^{-1}~Mpc^{-1}}$, $\Omega_\Lambda=0.73$ and
$\Omega_{\rm M}=0.27$ (Bennet et al. 2003).

\subsection{Flux-selected Spectroscopy}
In order to verify the presence of a direct correlation between $HR$ and 
count rate derived in $\S4$, we performed a flux-selected spectral analysis.
To this end we divided the 94 individual spectra into 5 bins according to
their average count rate (namely, $<0.25,~,0.25-0.30,~0.30-0.35,~0.35-0.40,
~{\rm and}~>0.40$ c/s). In each bin, the individual spectra were fitted
simultaneously with a PL model (absorbed by Galactic $N_{\rm H}$) with their
photon indices linked together and their respective normalizations free to
vary.

\begin{figure}
\begin{center}
\includegraphics[bb=45 32 355 300,clip=,angle=0,width=9cm]{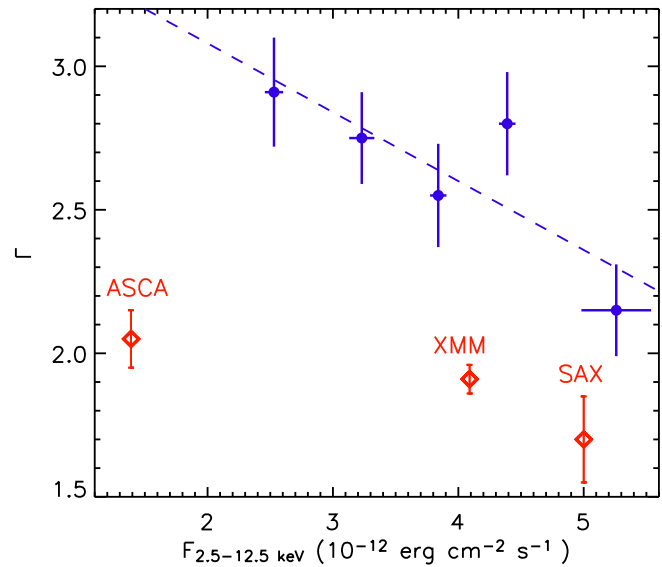}
\end{center}
\caption{$\Gamma$, obtained from the spectral analysis of flux-selected 
intervals, plotted against the average flux in the 2.5-12.5 keV band. 
The error-bars on $\Gamma$ are 1-$\sigma$ errors.
The dashed line represents the best linear fit $y=3.6\pm0.3-(0.24\pm0.08)x$.
For comparison, we have also plotted the
values corresponding to the \asca, \sax, and \xmm\ observations. To this end,
we have used {\tt PIMMS} to convert the observed flux into the \rxte\ energy
range,
assuming  the best-fit spectral parameters reported in Guainazzi et al.
(2003) and Gliozzi et al. (2004).}
\label{figure:fig8}
\end{figure}

The results are summarized in Table 2, where the reported errors on 
$\Gamma$ and flux are respectively 1$\sigma$ and  $\sigma/\sqrt N$, 
with $N$ being the number of individual spectra per bin. The upper limits
on EW correspond to the 90\% confidence level.
Table 2 indicates that the 2.5--12.5
keV spectra harden as the average flux increases. This is clearly shown in
Figure~\ref{figure:fig8}, where the values of $\Gamma$ for each count rate bin
have been plotted against their respective flux values. The dashed line 
represents the best linear fit, $y=3.6\pm0.3-(0.24\pm0.08)x$, which reveals
that the inverse correlation is significant at 3-$\sigma$ level.

For comparison, in Fig.~\ref{figure:fig8} we have also plotted the
values corresponding to the \asca, \sax, and \xmm\ observations. To this end,
we have used {\tt PIMMS} to convert the observed flux into the \rxte\ energy
range,
assuming  the best-fit spectral parameters reported in Guainazzi et al.
(2003) for \sax\ and \asca, and Gliozzi et al. (2004) for \xmm. The \asca, 
\sax, and \xmm\ photon
indices seem to follow the same inverse trend shown by \rxte, becoming flatter
 as the flux
increases. However, their values appear to be significantly smaller than those
obtained from the flux-selected \rxte\ spectral analysis. This apparent
discrepancy may be probably reconciled (at least for the \sax\ and \xmm\ 
data) by bearing in mind that: 1) Past studies have shown that photon
indices measured with the \rxte\ PCA are systematically steeper than those 
measured 
by other X-ray satellites (e.g., Yaqoob 2003). 2) The error-bars for the \rxte\
data in Fig.~\ref{figure:fig8} are at 1$\sigma$ level. On the other hand,
it appears more difficult to reconcile the \asca\ value with the extrapolation
of the \rxte\ best-fit trend, suggesting that the source was in a different
physical state (for example, it lacked a jet contribution).

We have also added a Gaussian line to the PL continuum (see Table 2 last 
column). Only for the bin with the lowest flux, the line is very marginally
significant (the value of $\chi^2$ decreases by 1.2 for one additional dof)
with EW$\sim$400 eV  and large uncertainties.
For bins with larger flux values, only upper limits on EW can be derived.
These findings indicate that a \feka\ line is never statistically required,
although its presence cannot be completely ruled out  when the source flux 
is at the lowest level measured by \rxte.

\section{Discussion} 

We have undertaken the first X-ray monitoring study of the FR~I galaxy \ngc,
investigating the temporal and spectral variability as well as time-averaged
and flux-selected spectral results.

By comparing the X-ray fluxes measured by different satellites over nearly
a decade of observations, it is clear that \ngc\ shows large flux changes
on long timescales.
For example, 
in October 1994 \asca\ measured a flux of $\sim 1.4\times 10^{-12}$ \flux\ 
in the 2--10 keV energy band,
whereas \sax\ observed a flux of $\sim 4.7\times 10^{-12}$ \flux\ in July 
2001, and \xmm\  of $\sim 4\times 10^{-12}$ \flux\ in March 2002. The high
throughput of the EPIC pn camera aboard \xmm\ also revealed the presence of
low-amplitude flux changes (of the order of $\sim$5\%) on timescales of a 
few ks.

The variability behavior detected in the current \rxte\ observations is in 
agreement with the
sparse evidence that had been gathered in the previous years, and show
conclusively that \ngc\ is a persistently variable source in X-rays.  The
frequent \rxte\ observations, spread over a period of one year, indicate 
that this
radio galaxy is characterized by persistent variability in the total 
(2.5--12 keV),
soft (2.5--5 keV), and hard (5--12 keV) energy bands. Throughout the 
observation, the flux appears to randomly change by a factor of $\sim$2
on timescales of a few weeks.

Previous X-ray studies, mostly based on the time-averaged spectral 
analysis, have revealed a dual behavior  for \ngc: sometimes the source 
appears to behave like a Seyfert galaxy with a strong \feka\ line (e.g.,
Turner et al. 1997; Sambruna et al. 1999; Gliozzi et al. 2004), whereas in 
other occasions \ngc\ seems to be more consistent with a blazar-like 
behavior, showing a featureless X-ray spectrum (e.g., Guainazzi et al. 2003).
The densely
sampled, year-long \rxte\  observations and the investigation of the flux and
spectral variability properties offer an alternative and model-independent
way to shed some light on this issue.

\subsection{Evidence from the flux variability properties}

Unlike the brightest blazars frequently monitored by
\rxte\ (e.g., Kataoka et al. 2001, Cui 2004, Xue \& Cui 2005), \ngc\ does 
not show any prominent flare on any observable timescale, nor does it show
the large amplitude variability typically observed in several Seyfert-like 
objects during yearly-long monitoring campaigns (e.g., Markowitz \& Edelson 
2001).

On one hand, the apparent inconsistency with the large variability 
observed in 
 Seyfert galaxies can be  explained by the lower values of black hole mass
in the latter objects, which is typically  one order of magnitude lower than 
\ngc. On the other hand, the lack of prominent flares in the NGC6251 light 
curve, which instead
characterize the blazar light curves, can be understood by keeping in mind
that the blazars monitored by RXTE are the brightest members of
this AGN class and  that the observations are often triggered only during 
their flaring activity. Nonetheless, blazar monitoring campaigns with
baselines covering several years reveal that also the brightest blazars
alternate prominent flaring activity with  ``quiescent periods'' that are
characterized by moderate flux variations.
Indeed, prolonged periods of moderate variability have been detected in several
blazars (e.g., B\"ottcher et al. 2005; Marscher 2006).

A model-independent study of the \rxte\ variability properties of the 
prototypical blazar Mrk~501, similar to the one performed on \ngc\ in this 
work, offers the best opportunity for a more quantitative comparison 
of \ngc\ with a typical blazar behavior (Gliozzi et al. 2006).
For instance, Mrk 501, which showed a large outburst in 1997, underwent a 
progressive decrease of
its activity in the following years, resulting in a lower mean count rate
accompanied by lower variability. Specifically, in 1999, 
when Mrk 501 reached a minimal flux value, the light curve was characterized by
variations of the order of $\sim$2 on timescales of 20--30 days, which are 
fully consistent with those detected in \ngc.

Before proceeding further, it must be pointed out that the blazars regularly
monitored over the past years with \rxte\ (including Mrk~501) are basically 
all High-peaked blazars
(HBLs), with the synchrotron component peaking in the X-ray range and a
second spectral component, generally attributed to inverse Compton scattering,
peaking
at TeV energies. On the other hand, in the blazar framework, the broadband SED
of \ngc\ is consistent Low-peaked blazars (LBLs), with the inverse Compton
component peaking in the X-rays (Chiaberge et al. 2003; Guainazzi et al. 2003).
As a consequence, a formal comparison of the X-ray properties of \ngc\ should
be in principle performed using the TeV properties of Mrk~501. However, 
detailed studies
of X-ray and TeV emissions in HBLs have demonstrated the existence of a tight
correlation between these energy bands, indicating that the X-ray and TeV radiation follow the same variability trend (e.g., Fossati et al. 2004; Gliozzi 
et al. 2006). Further support to this conclusion comes from recent 
investigations of the TeV properties of Mrk~501 and Mrk~421 carried out
with the MAGIC and Whipple telescopes (Blazejowki et al. 2006; 
Albert et al. 2007). These studies demonstrate that also in the $\gamma$-ray
energy band HBLs show the typical blazar spectral variability behavior 
observed in X-rays, that is a spectral hardening when the 
source brightens and a fractional variability more pronounced at
higher energies (e.g., Zhang et al. 1999; Fossati et al. 2000; Gliozzi et al.
2006; Rebillot et al. 2006 and references therein). We therefore conclude
that it is appropriate to compare the X-ray properties of Mrk~501 with those
of \ngc.

Another relevant finding from the flux variability analysis of \ngc\ is that 
the fractional variability appears to be more pronounced in the hard than
in the soft energy band: $\Delta F_{\rm var}\equiv F_{\rm var,hard}-
F_{\rm var,soft}=(10\pm6)\%$. This result can be quantitatively compared
with similar studies carried out on Seyfert galaxies monitored with \rxte\
for several months or years (Markowitz \& Edelson 2004). Specifically, all 
the Seyfert galaxies with long \rxte\ monitoring yield negative values of
$\Delta F_{\rm var}$ ranging between -3.5 and -24.8, with a mean
value of $-10.3\pm1.5$.  On the other hand, a similar study, carried out
using \rxte\ monitoring data of the blazar Mrk~501 between 1997 and 2000, 
yielded positive values of $\Delta F_{\rm var}$ ranging between 5 and 
42, with a mean value of $15\pm1$.

We can therefore conclude that the limited amplitude of the \ngc\ flux
variability can be explained equally well in both Seyfert-like and 
blazar-like scenarios. However, the fractional variability behavior 
is inconsistent with the trend generally observed in Seyfert galaxies,
but fully consistent with the typical blazar behavior observed in Mrk~501.

\subsection{Evidence from the spectral properties} 
A comparative analysis of the
spectral variability of \ngc\ and the typical behaviors observed
in both radio-quiet AGNs (where the X-rays are thought to be produced by 
Comptonization in the corona that is closely connected with the accretion 
disk) and radio-loud jet-dominated AGNs (whose radiation over the entire 
energy range is ascribed to jet emission) can help us understand whether
the X-ray radiation from \ngc\ is dominated by jet or accretion-related 
emission.

A flux-selected spectral analysis (in agreement with
the $HR-ct$ plot) has shown that the X-ray spectrum of \ngc\
hardens as the source brightens, following the linear correlation
$\Gamma\propto -0.24(\pm0.08)\times F_{\rm X}$.  A direct comparison of
this result with a similar spectral study, carried out on 4 Seyfert galaxies 
monitored with \rxte, indicate that the latter have always  positive slopes
in the $\Gamma-F_{\rm X}$ plane ranging between 0.05 and 0.15 with  a
mean value of $0.09\pm0.01$ (Papadakis et al. 2002). As a consequence,
the spectral variability behavior of \ngc\ appears to 
be inconsistent
with the typical Seyfert-like trend. On the other hand,
the  $\Gamma-F_{\rm X}$ slope measured for Mrk~501 during the weakly variable 
period of 2000,
$-0.37\pm0.06$, appears to be fully consistent with the behavior of \ngc\
during the \rxte\ monitoring campaign.

The existence of a Seyfert-like component suggested by previous X-ray studies
was essentially based on the presence of a strong \feka\ line. Unfortunately,
the low S/N spectra obtained during the \rxte\ monitoring coupled with the
poor spectral resolution of the \rxte\ PCA hampers a detailed investigation
of this issue. Indeed, if the 2-month averaged PCA spectra are fitted with
a model including a power law and a Gaussian line with spectral parameters
fixed at the best-fit values obtained during the \asca\ observation (we
conservatively assumed the best-fit parameters reported in Guainazzi et al.
2003), the results are statistically 
indistinguishable from the fits obtained
using a simple power law. This indicates that \rxte\ is unable to confirm or
refute the presence of a \feka\ line, and suggests that only a relatively long
exposure of \xmm\
with its superior capabilities is able to detect the presence of the
line when the average flux of \ngc\ is relatively high.

\subsection{The origin of X-rays in \ngc}
The primary goal of this work is to investigate the origin of the X-rays
in \ngc\ and in particular to assess the role played by the putative jet.
At first sight, the possibility that the jet may dominate the radiative
output of a radio galaxy  may be surprising, and even more so for NGC6251,
which has a Mpc radio jet (and hence appears forming a large viewing angle) 
and Seyfert-like emission during the low flux state.
However, under specific circumstances, the jet-dominance hypothesis becomes 
plausible. This is the case when the base of the jet is not well collimated 
and  the X-rays are produced by a part of the outflowing material that points 
towards the observer. Indeed, this is the framework proposed to explain the 
TeV  emission detected in another radio galaxy M87 (Aharonian et al. 2006). 
Alternatively, if the 
base of the jet is tilted with respect to the large scale jet
and forms a small viewing angle, the possible dominance of the jet in the 
2--12 keV range can be naturally explained. In fact, this is the 
scenario put forward  by Jones \& Wehrle (2002)  to explain 
the large jet/counterjet brightness ratio on parsec scales inferred for
\ngc\ from VLBA observations.

By comparing the spectral variability properties of \ngc\ with those of 
radio-quiet 
AGNs and blazar objects, we find that they are certainly inconsistent with
a typical Seyfert-like behavior but 
fully consistent with blazars. 
Combining pieces of information from the model-independent analysis with the
findings from the flux-selected spectral analysis, we are led to the conclusion
that, during the \rxte\ monitoring campaign, the bulk of the hard X-ray
radiation from \ngc\ was dominated by the emission from the unresolved base of
the jet. Nonetheless, the presence of a disk-corona component, detected in 
previous observations with \asca\ and \xmm, cannot be ruled out: the upper
limits measured on the equivalent width of the \feka\ line are indeed fully
consistent with the values measured by \asca\ and \xmm, but
the low S/N of the spectra hampers a more quantitative analysis.

In conclusion, we can try to exploit the main results from this work 
(i.e., the jet
dominance in the high flux state, with a possible contribution from
a Seyfert-like component emerging
at low flux values) to derive some constraints on the accretion process
at work in \ngc. Assuming that an accretion-related component is
always present in \ngc\ and dominates during the low flux state, we can use
the average flux measured in the low count rate bin (see Table 2) to compare
the corresponding 2--10 keV luminosity -- $L_{\rm X}=6.6\times 10^{42}$ \lum --
to the Eddington value readily derived from the black hole mass estimate.
The relatively high value derived, $L_{\rm X}/L_{\rm Edd} > 5\times 10^{-5}$,
confirms that \ngc\ is a bright FR~I galaxy that is close to the FR~I/FR~II
dividing line in terms of power of the central engine. However, it is not
possible to derive any firm conclusion on the nature of the accretion process
given the unknown contribution of the jet in the low-flux state.

\section{Summary and Conclusions} 

We have used data from a year-long \rxte\ monitoring campaign to study the 
spectral variability of \ngc\, following model-independent and spectral model 
fitting methods. The main results can be summarized as follows:

\begin{itemize}

\item Throughout the monitoring campaign and especially during the last 4 
months, \ngc\ was in a relatively high-flux state, with values of the 
2--10 keV absorbed flux comparable to that observed by \sax\ in 2001.

\item The light curves show persistent 
 variations by a factor of 2 on timescales weeks/months in the total
(2.5--12 keV), soft (2.5--5 keV), and hard (5--12 keV) energy bands.

\item The fractional variability, computed over the soft and hard energy
bands, reveals an enhanced variability at harder energies ($F_{\rm var}
\sim$20\%) compared to the soft band ($F_{\rm var}\sim$10\%).

\item There is evidence of a positive trend in the $HR-ct$ plot (or,
analogously, of a negative trend in the $\Gamma$-flux plot); in other
words the spectrum hardens as the flux increases.

\item The 2-month averaged spectra are well fitted by a power-law model, 
with $\langle\Gamma\rangle\simeq2.5$ and no indication for an \feka\ 
line. Combining all the 2-month averaged spectra yielded EW$<$ 104 eV.
Only for the lowest flux-selected spectrum, there is marginal 
evidence for a \feka\ line. However, the low S/N does not allows one to
put reliable constraints on the putative Seyfert-like component.
\end{itemize}

The inconsistency of the spectral variability behavior of \ngc\
with the typical trend observed in Seyfert galaxies and the 
similarity with blazars lead support to a jet-dominated scenario.
However, based on the \rxte\ observations,
a substantial contribution from a disk-corona system cannot be ruled out.

\begin{acknowledgements}
We thank the referee for the
comments and suggestions that improved the clarity of the paper.
MG acknowledges support by the RXTE Guest Investigator Program
under NASA grant 200858. Funds from the NASA LTSA grant
NAG5-10708 are also gratefully acknowledged.
\end{acknowledgements}

\begin{table} 
\caption{Time-resolved Spectral results}
\begin{center}
\begin{tabular}{lcccc} 
\hline        
\hline
\noalign{\smallskip}        
Interval &  $\chi^2_{\rm red}$ & $F_{\rm 2-10~keV}$                 & $\Gamma$                 & EW$_{\rm FeK\alpha}$ \\
\noalign{\smallskip}
         &                        & $(10^{-12}{\rm erg~cm^{-2}~s^{-1}})$ &                        &  (eV)                \\
\noalign{\smallskip}
\hline  
\noalign{\smallskip}
1       & 0.60               &  4.1                                & $2.6_{-0.4}^{+0.4}$ &  $< 366$   \\
\noalign{\smallskip}
\hline
2       & 0.53               &  4.5                                & $2.6_{-0.3}^{+0.3}$ &  $< 466$   \\
\noalign{\smallskip}
\hline
3       & 0.45               &  4.4                                & $2.5_{-0.4}^{+0.3}$ &  $< 444$   \\
\noalign{\smallskip}
\hline
4       & 0.81               &  4.0                                & $2.4_{-0.4}^{+0.4}$ &  $< 756$   \\
\noalign{\smallskip}
\hline
5       & 1.00               &  5.2                                & $2.7_{-0.3}^{+0.3}$ &  $< 181$   \\
\noalign{\smallskip}
\hline
6       & 0.59               &  4.8                                & $2.4_{-0.3}^{+0.3}$ &  $< 246$   \\
\noalign{\smallskip}
\hline

\end{tabular}
\end{center}
{\bf Note:}  All errors and upper limits refer to 90\% confidence levels.
\label{tab1}
\end{table}

\begin{table} 
\caption{Flux-selected Spectral results}
\begin{center}
\begin{tabular}{cccccc} 
\hline        
\hline
\noalign{\smallskip}        
Individual Spectra & Count Rate      &  $\chi^2_{\rm red}$ & $F_{\rm 2.5-12.5~keV}$                  & $\Gamma$      & EW$_{\rm FeK\alpha}$ \\
\noalign{\smallskip}
                   & $({\rm s^{-1}})$&                      & $(10^{-12}{\rm erg~cm^{-2}~s^{-1}})$ &             &  (eV)                \\
\noalign{\smallskip}
\hline  
\noalign{\smallskip}
31 & $< 0.25$       & 0.53            &  $2.5\pm0.1$   & $2.9\pm0.2$ &  $<903$   \\
\noalign{\smallskip}
\hline
26 & $0.25-0.30$   & 0.55            &  $3.2\pm0.1$   & $2.7\pm0.2$ &  $<540$   \\
\noalign{\smallskip}
\hline
16 & $0.30-0.35$   & 0.50            &  $3.8\pm0.1$   & $2.5\pm0.2$ &  $<205$   \\
\noalign{\smallskip}
\hline
12 & $0.35-0.40$   & 0.59            &  $4.4\pm0.1$   & $2.8\pm0.2$ &  $<378$   \\
\noalign{\smallskip}
\hline
9 & $> 0.40$       & 0.60            &  $5.3\pm0.3$   & $2.2\pm0.2$ &  $<305$   \\
\noalign{\smallskip}
\hline
\end{tabular}
\end{center}
{\bf Note:} The errors on $\Gamma$ are 1$\sigma$ (68\% confidence level), whereas the upper limits 
on EW are 90\% confidence levels.
\label{tab2}
\end{table}

\end{document}